# Flexible frontiers for text division into rows

**Dan L. Lacrămă[1], Ioan Şnep[2]**
[1] ”Tibiscus” University of Timişoara, Romania
[2] NMS Institute, Paris, France

**ABSTRACT**: This paper presents an original solution for flexible hand-written text division into rows. Unlike the standard procedure, the proposed method avoids the isolated characters extensions amputation and reduces the recognition error rate in the final stage.
**KEYWORDS:** pattern recognition, image processing

## Introduction

The standard methodology for any automatic reading systems contains a multiple segmentation procedure before the character recognition stage. The text body is first divided into rows, than each word inside the row is detached and finally word's letters are isolated. Sometimes, this decomposition is continued, and even the characters are cleaved in their sub-components in order to detect lines loops and breaks in their structure [SGS93].

The cutting of the text body into rows is traditionally done by straight line borders, and this proves to be highly efficient for printed text samples as it is shown in Figure 1.1.a. This procedure is also adequate if the text is carefully hand-written in order to avoid any partial superposition between the descendant zone of the upper row and the ascendant of its lower neighbour.

Unfortunately this is not the case for the most of the hand-written samples and consequently there is no chance for the frontier to bypass one or more of the ascending or descending extensions present in the inter-row region as it can be seen in the Figure 1.1.b.

These amputations usually do not affect the recognition of the cleaved character but create a great error hazard in the classification of the neighbour to which the cut "tail" is appended. For example the little part cleaved from the letter 'p' in word "pt." is likely to cause the wrong classification of the first loop of the 'm' in the word "imagini" as an 'i' and consequently a non-word label (recognition fail) for the resulting "iinagini" at the end of the reading [Gil96]. This peril is bigger when texts are written in a language where the alphabet contains letters with accents or other isolated upper or lower extensions (e.g. Romanian specific characters 'ă', 'â' 'ş', and 'ţ').

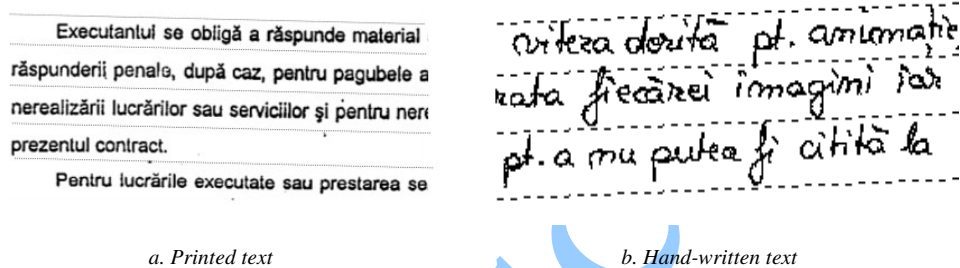

a. Printed text                    b. Hand-written text

**Figure 1.1.** *Straight line borders text segmentation into rows*

Bottom edge following borders segmentation is a second methodology recommended in the literature. The basic idea is the construction of frontiers point by point nearby the base edge of the words. The link between two consecutive words is done usually by a straight line. After this first construction step, the border is redrawn with straight line segments in order to reduce its complexity (i.e. contour relaxation) [LF95].

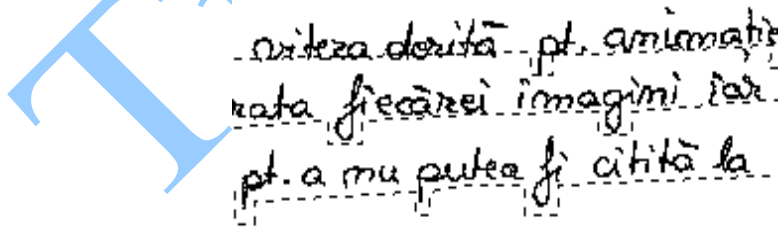

**Figure 1.2.** *Bottom edge following borders method for text segmentation into rows*

The algorithm performs a very good segmentation into rows for texts in English but constantly amputates the descending isolated parts (i.e. diacritic signs) of special characters in other national languages alphabets. This problem is illustrated in Figure 1.2. where the word "animaţie" letter 'ţ' low part is cut off from the upper row and will be appended to the lower one causing latter a great amount of problems in the recognition of character 'r' from "iar".

Another disadvantage of this procedure is the fact that it is time consuming because it does the edge following point by point all over the given document and this affect the speed for all the segmentation process. The contour relaxation in the end solve the problem of simplifying the text image division into individual rows images, but it also need some extra time to be done.

Therefore this paper propose a technique able to avoid both the amputation of the linked and isolated ascending or descending extensions of the characters in the process of row segmentation of the text body.

## 1 Flexible frontiers segmentation method

The flexible borders are usually straight line segments which link together short edge following curve segments existing only in the neighbourhood of a character - frontier intersection point. Hence the row dividing border contains:

* line segments extending until a letter edge is reached
* contour following curves extending until a new line segment can be started
*

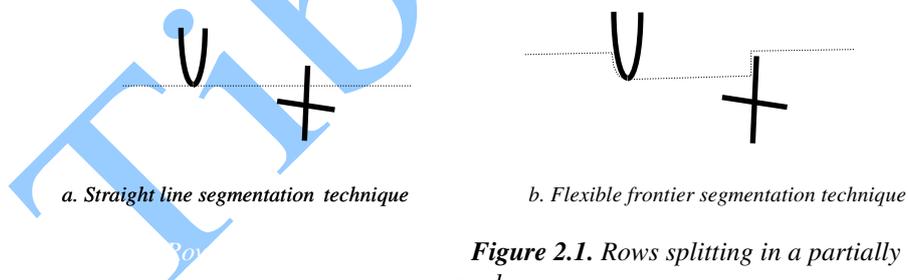

a. Straight line segmentation technique    b. Flexible frontier segmentation technique

**Figure 2.1.** Rows splitting in a partially superposed zone

The contour pursuing method used for the algorithm implementation was the "Hand on the wall following" standard procedure described in [GW93] and it proved to be an efficient and low time consuming. A comparison between the results of the classic straight line border method and the method proposed by this paper is given in Figure 2.1 for a case that is quite frequent in hand-written texts. The Logic flow chart of the flexible border procedure routine is given in Figure 2.2.

Because it was very important to avoid amputation of isolated character's parts, every time when such an extension is met the border detour realised by the pursuing curve was orientated firstly to the exterior of the row. Only if it arrives to the central part of the adjacent row the procedure is taken from its begging with a detour to the interior. Figure 2.3 illustrates this kind of behaviour and its importance for a correct separation of the rows in a Romanian text.

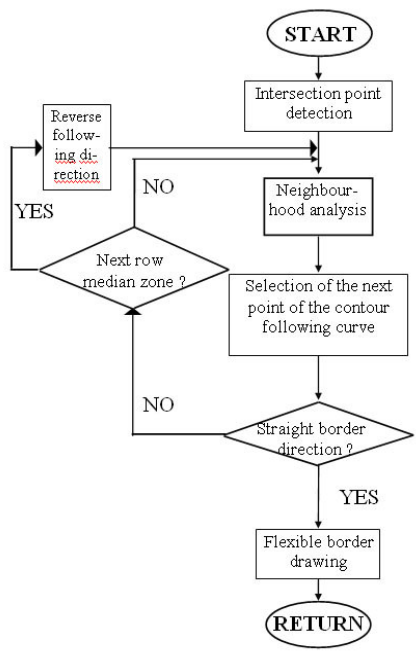

*Figure 2.2. Logic flow chart of the flexible border procedure routine*

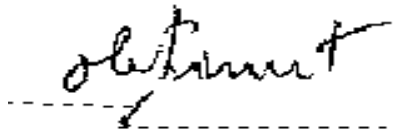

*Figure 2.3. Avoiding amputation with exterior detour*

It has to be stressed here that every straight line segment of the flexible border has the same direction given by the medium row angle. This is necessary because usually hand-written rows are not exactly parallel with the document image horizontal. This is due both to the human error and the impossibility of exact alignment in the acquisition process [LSO95].

The routine waits for an intersection point to occur and starts the contour tracking by looking for a neighbouring pixel that is a background - letter edge pixel. When such a pixel is found, it is declared point of the frontier and a new border pixel is searched in its own neighbourhood.

First the contour following direction is oriented to the outside of the row. The "Next row median zone?" if block controls the frontier in order to avoid its entrance in the middle zone of the nearby row. If this happens the direction is reversed and a new tracking procedure is started from the beginning intersection point.

This point by point construction, similar to the one in the second method described in the paragraph above, ends in the point where a line segment with the medium row angle can be started. This point is detected in the "Straight border direction?" if block. If it encounters another intersection the pursuing is repeated with the same algorithm, if not the line goes until the right end of the image. During each curved segment the border is drawn by the above routine, during line segment the border is established and designed by the main programme.

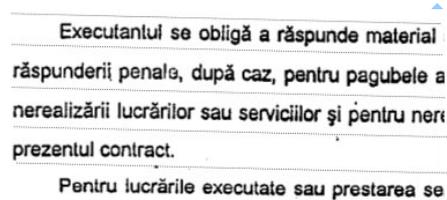
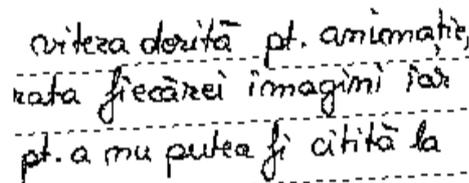

*a. Printed text*                    *b. Hand-writting text*

**Figure 3.1.** *Flexible frontiers segmentation results*

## 2 Experimental results

The Figure 3.1.a. and b. show the test of the Flexible frontier segmentation technique over the same samples as the two classic methods presented in the first paragraph. There are no errors and the text body is correctly cut into rows in both cases. This remarkable adaptability to quite different situations is an important and useful feature of the proposed procedure.

Extending experiments to a set of 24 test samples it was proved that the method was able to reduce segmentation errors with: 63.7% in comparison with the Straight line border method and 12.2% in comparison with the Bottom edge following borders method.

Another important gain of the proposed procedure is the fact that it preserves the speed of the straight line border method in samples where no partial superposition occurs (sample from Figure 3.1. is segmented in practically the same time with both straight and flexible frontiers method). This is possible because the contour following procedure is a routine becoming active only when it is needed, in the moment of meting a border - character intersection.